\def\bbbq{{\mathchoice 
{\setbox0=\hbox {$\displaystyle\rm Q$}\hbox
{\raise0.15\ht0\hbox to0pt{\kern0.4\wd0\vrule height0.8\ht0\hss}\box0}}
{\setbox0=\hbox {$\textstyle\rm Q$}\hbox
{\raise0.15\ht0\hbox to0pt{\kern0.4\wd0\vrule height0.8\ht0\hss}\box0}}
{\setbox0=\hbox {$\scriptstyle\rm Q$}\hbox
{\raise0.15\ht0\hbox to0pt{\kern0.4\wd0\vrule height0.7\ht0\hss}\box0}}
{\setbox0=\hbox {$\scriptscriptstyle\rm Q$}\hbox
{\raise0.15\ht0\hbox to0pt{\kern0.4\wd0\vrule height0.7\ht0\hss}\box0}}
}}

\def\bbbr{{\rm I \!\! R}} 
 
\def\bbbc{{\mathchoice 
{\setbox0=\hbox {$\displaystyle\rm C$}\hbox
{\hbox to0pt{\kern0.4\wd0\vrule height0.9\ht0\hss}\box0}}
{\setbox0=\hbox {$\textstyle\rm C$}\hbox
{\hbox to0pt{\kern0.4\wd0\vrule height0.9\ht0\hss}\box0}}
{\setbox0=\hbox {$\scriptstyle\rm C$}\hbox
{\hbox to0pt{\kern0.4\wd0\vrule height0.9\ht0\hss}\box0}}
{\setbox0=\hbox {$\scriptscriptstyle\rm C$}\hbox
{\hbox to0pt{\kern0.4\wd0\vrule height0.9\ht0\hss}\box0}}
}}

\font\fivesans=cmss10 at 5pt 
\font\sevensans=cmss10 at 7pt 
\font\tensans=cmss10   
\newfam\sansfam  
\textfont\sansfam=\tensans\scriptfont\sansfam=\sevensans  
\scriptscriptfont\sansfam=\fivesans  
\def\sans{\fam\sansfam\tensans} 
\def\bbbz{{\mathchoice {\hbox{$\sans\textstyle Z\kern-0.4em Z$}}  
{\hbox{$\sans\textstyle Z\kern-0.4em Z$}}  
{\hbox{$\sans\scriptstyle Z\kern-0.3em Z$}}  
{\hbox{$\sans\scriptscriptstyle Z\kern-0.2em Z$}}}} 
   
\def\bbbh{{\rm I \!\! H }}

\def\slash#1{#1\kern-0.65em /}
\def\dirac{{\raise0.09em\hbox{/}}\kern-0.58em\partial}
\def\Dirac{{\raise0.09em\hbox{/}}\kern-0.69em D}

\def\tr{{\rm Tr}}

\def\m@th{\mathsurround=0pt}
\def\eqalign#1{\null\,\vcenter{\openup 3pt \m@th
\ialign{\strut\hfil$\displaystyle{##}$&$\displaystyle{{}##}$\hfil
\crcr#1\crcr}}\,}

\documentstyle[12pt]{article}
\begin{document}

\title{Deformations of Differential Calculi}

\author{J. Madore \\
        Laboratoire de Physique Th\'eorique et Hautes 
        Energies\thanks{Laboratoire associ\'e au CNRS, {\rm URA D0063}}\\
        Universit\'e de Paris-Sud, B\^at. 211, F-91405 Orsay
\and    J. Mourad \\
        GPS, Universit\'e de Cergy Pontoise\\
        Site de St. Martin, F-95302 Cergy Pontoise
\and    A. Sitarz\\
        Johannes Gutenberg Universit\"at\\
        Institut f\"ur Physik, D-55099 Mainz    
       }
\date{} 
\maketitle

\abstract{
It has been suggested that quantum fluctuations of the gravitational
field could give rise in the lowest approximation to an effective
noncommutative version of Kaluza-Klein theory which has as extra hidden
structure a noncommutative geometry.  It would seem however from the
Standard Model, at least as far as the weak interactions are concerned,
that a double-sheeted structure is the phenomenologically appropriate
one at present accelerator energies.  We examine here to what extent
this latter structure can be considered as a singular limit of the
former.
}

\vfill
\noindent
LPTHE Orsay 95/75
\medskip
\noindent
November, 1995
\bigskip
\eject

\section{Motivation}

It has been suggested (Madore \& Mourad 1995) that quantum fluctuations
of the gravitational field (Deser 1957, Isham {\it et al.} 1971) could
give rise in the lowest approximation to an effective noncommutative
version (Madore, 1990) of Kaluza-Klein theory which has as extra hidden
structure a noncommutative geometry described by, for example an algebra
$M_n$ of $n \times n$ complex matrices (Dubois-Violette {\it et al.}
1989, 1991). From the Standard Model it would seem however, at least as
far as the weak interactions are concerned, that the phenomenologically
most appropriate extra structure at accelerator energies would be one
based on the algebra $\bbbh \times \bbbc$ (Connes \& Lott 1992, Iochum
\& Sch\"ucker 1994). It would be possible to reconcile the implications
of the quantum fluctuations with the experimental facts if one could
show that in a consistent natural way one of the former structures could
be deformed into the latter. We shall study here a simpler problem; we
shall consider the algebra $\bbbc \times \bbbc$ as a singular
contraction of the algebra $M_2$.

The algebra $\bbbc \times \bbbc$ is the algebra of complex-valued
functions on the space of two points. The algebra $M_2$ can be also
considered as describing the two-point structure, but in a symmetric
way. The adjoint action of the group $SO_3$ on $M_2$ is the analogue of
rotational symmetry in ordinary geometry. The contraction does not
respect this symmetry; it is obviously broken by a specific choice of
two points.  One can imagine the algebra $M_2$ as describing a very
fuzzy sphere with only two distinct regions. One can consider the
contraction as a singular deformation of the sphere; its effect is to
squeeze the sphere to a line and the two regions onto the end points.
The algebra $\bbbc \times \bbbc$ can be also considered as describing a
classical spin system which can take only 2 values. The corresponding
quantum spin system is described by the algebra $M_2$ (Bratteli \&
Robinson 1989). So $\bbbc \times \bbbc$ can be considered as a limit of
$M_2$ when $\hbar \rightarrow 0$.

The 2-point model and the $SO_3$-symmetric model depend each on a
differential calculus over the corresponding algebra.  It is our purpose
here to show that the contraction can be extended to a map between the
corresponding differential calculi.  In the next two sections we recall
briefly the two models and in Section~4 we define the contraction. In
Section~5 we present our conclusions.

\section{The noncommutative algebra}

Let $\lambda_a$, for $1 \leq a \leq 3$, be an anti-hermitian basis of
the Lie algebra of the special linear group $SL_2$.  The product
$\lambda_a \lambda_b$ can be written in the form
$$
\lambda_a \lambda_b = {1\over 2} C^c{}_{ab} \lambda_c - 
                      {1\over 2} g_{ab}.                           \eqno(2.1)
$$
The structure constants $C^c{}_{ab}$ can be chosen real and the $g_{ab}$ 
can be expressed in terms of them by the equation
$$
g_{ab} = - {1\over 4}C^c{}_{ad}C^d{}_{bc}.
$$
We shall lower and raise indices with $g_{ab}$ and its inverse $g^{ab}$.
The tensor $C_{abc}$ is completely antisymmetric. The $\lambda_a$ can be
chosen such that the $g_{ab}$ are the components of the euclidean
metric in 3 dimensions. For example they can be written in terms of
the Pauli matrices $\sigma_a$ as $\lambda_a = - (i/ \sqrt 2) \sigma_a$.
The structure constants are given then by $C_{123} = \sqrt 2$.  The
matrix algebra $M_2$ is generated by the $\lambda_a$ as an algebra.

The derivations
$$
e_a = m\, {\rm ad}\,\lambda_a                                       \eqno(2.2)
$$
form a basis over the complex numbers for the set ${\rm Der}(M_2)$ of
all derivations of $M_2$.  The mass parameter $m$ has been included so
that they have the correct physical dimensions.  We recall
that the adjoint action is defined on the element $f \in M_2$ by 
$e_a f = m\, {\rm ad}\,\lambda_a (f) = m\,[\lambda_a, f]$.  Any element 
$X$ of ${\rm Der}(M_2)$ can be written as a linear combination 
$X = X^a e_a$ of the $e_a$ where the $X^a$ are complex numbers. The
3-dimensional vector space ${\rm Der}(M_2)$ is a Lie-algebra, the Lie
algebra of the group $SU_2$; it is the analogue of the Lie algebra of
global vector fields on a smooth manifold. In particular the derivations
$e_a$ satisfy the commutation relations 
$[e_a, e_b] = m \, C^c{}_{ab} \, e_c$.

We define the 1-forms $\Omega^1(M_2)$ over $M_2$ just as one does in the
commutative case (Dubois-Violette 1988). We define $df$ for $f \in M_2$ 
by the formula
$$
df(e_a) = e_a f.                                                   \eqno(2.3)
$$
This means in particular that
$$
d\lambda^a(e_b) = m\, [\lambda_b, \lambda^a ] = - m\, C^a{}_{bc}\lambda^c.
$$
The set of $d\lambda^a$ constitutes a system of generators of
$\Omega^1(M_2)$ as a left or right module but it is not a convenient
one. For example $\lambda^a d\lambda^b \neq d\lambda^b \lambda^a$. There
is a better system of generators $\theta^a$ completely characterized 
by the equations
$$
\theta^a(e_b) = \delta^a_b.                                        \eqno(2.4)
$$
The $\theta^a$ are related to the $d\lambda^a$ by the equations
$$
d\lambda^a =  m \, C^a{}_{bc}\, \lambda^b \theta^c, \qquad
\theta^a = m^{-1} \lambda_b \lambda^a d\lambda^b.                  \eqno(2.5)
$$
Because of the relation (2.4) we have
$$
\theta^a \, \theta^b = - \theta^b \, \theta^a, \qquad
\lambda^a \theta^b = \theta^b \lambda^a.
$$
The products here are defined by
$$
(\theta^a \, \theta^b) (e_c, e_d) = {1\over 2} \delta^{ab}_{cd}, \qquad      
(\lambda^a \theta^b) (e_c) = \lambda^a \theta^b (e_c),           \qquad 
(\theta^b \lambda^a) (e_c) = \theta^b (e_c) \lambda^a.              \eqno(2.6)
$$
The $\theta^a$ generate an exterior algebra $\bigwedge^*$ of dimension 
$2^3$ and they satisfy the same structure equations as the components of
the Maurer-Cartan form on the special linear group $SL_2$:
$$
d\theta^a = -{1 \over 2} m\, C^a{}_{bc} \, \theta^b \, \theta^c.    \eqno(2.7)
$$
Using these relations it is easy to see that the algebra 
$\Omega^*(M_2)$ is equal to the tensor product of $M_2$ and
$\bigwedge^*$:
$$
\Omega^*(M_2) = M_2 \otimes_\bbbc {\textstyle \bigwedge}^*.         \eqno(2.8)
$$
It is therefore of dimension $4 \times 2^3$.

The homology of the complex $(\Omega^*(M_2), d)$ satisfies the
isomorphisms
$$
H^2(M_2) \simeq H^1(M_2) = 0,            \quad
H^3(M_2) \simeq H^0(M_2) = \bbbz,        \quad
H^p(M_2) = 0, \; p \geq 4.                                          \eqno(2.9)
$$
In the absence of a possible definition of homology groups this can be
considered as a form of Poincar\'e duality. An arbitrary topological
space $V$ has homology $H_*(V)$ as well as cohomology $H^*(V)$ which, if
$V$ is a smooth manifold of dimension $n$, are isomorphic: 
$H_*(V) \simeq H^{n-*}(V)$. Using the (co)homology Chern characters this
isomorphism can be expressed in terms of the algebra ${\cal C}(V)$ of
smooth functions on $V$ as an isomorphism of $K^*\big({\cal C}(V)\big)$
onto $K_*\big({\cal C}(V)\big)$ and as such generalized to a property of
arbitrary algebras.  Connes (1995) has stressed the importance of the
role of this version of Poincar\'e duality as a necessary condition in
distinguishing noncommutative geometries which can be considered as
`smooth' from those which are only `topological'. By `smooth' we mean
here something stronger, an algebra whose differential calculus is 
based on derivations.

From the generators $\theta^a$ we can construct the 1-form
$$
\theta = - m \lambda_a \theta^a 
       = - {1\over 2} \lambda_a d\lambda^a 
       =   {1\over 2} d\lambda_a \lambda^a.                        \eqno(2.10)
$$
It follows directly from the definitions that the exterior derivative
$df$ of an element $f \in M_2$ can be written in terms of a commutator 
with $\theta$:
$$
df = - [\theta,f].                                                 \eqno(2.11)
$$
This is not true however for an arbitrary element of $\Omega^*(M_2)$.
From (2.5) and (2.7) it follows that
$$
d \theta + \theta^2 = 0.                                           \eqno(2.12)
$$
As a left or right $M_2$-module $\Omega^1(M_2)$ is free with three
generators but from (2.11) one sees that as a $M_2$-bimodule
$\Omega^1(M_2)$ is generated by $\theta$ alone.

It is interesting to note that the differential algebra $\Omega^*(M_2)$
can be imbedded in a larger algebra in which there is an element
$\theta$ such that (2.11) is satisfied for all elements of
$\Omega^*(M_2)$. For the details we refer to Madore (1995). In general
any differential calculus $\Omega^*({\cal A})$ over an arbitrary
associative algebra ${\cal A}$ can be enlarged by addition of an element
$\theta$ such that (2.11) is satisfied for all elements of
$\Omega^*({\cal A})$. In the case of the de~Rham calculus over a smooth
manifold the extra element can be chosen to be the phase of the Dirac
operator (Connes 1994).

One defines a Yang-Mills potential to be an anti-hermitian element
$\omega \in \Omega^1(M_2)$. We can write it using $\theta$ as
$$
\omega = \theta + \phi.                                           \eqno(2.13)
$$
The unitary elements $U_2$ of the algebra $M_2$ can be considered as the
group of gauge transformations. For $g \in U_2$ we have
$\omega \mapsto g^{-1} \omega g + g^{-1} dg$.  It is clear that in
particular $\theta \mapsto \theta$ and therefore that 
$\phi \mapsto g^{-1} \phi g$. Expand $\phi$ in terms of the basis
$\theta^a$: $\phi = \phi_a \theta^a$. It follows from (2.12) that the
field strength $F$ of the potential $\omega$ is given by
$$
F = d\omega + \omega^2 = {1\over 2} F_{ab} \theta^a \theta^b, \qquad   
F_{ab} = [\phi_a, \phi_b] - m \, C^c{}_{ab} \phi_c.               \eqno(2.14) 
$$

Let ${\cal C}(\bbbr^4)$ be the algebra of smooth functions on
space-time.  Using $\Omega^*(M_2)$ one can construct a differential
calculus over the algebra ${\cal C}(\bbbr^4) \otimes M_2$ which
describes a fuzzy version of space-time.  The electromagnetic lagrangian
in the extended space can be written
$$
{\cal L} = {1\over 4} \tr (F_{\alpha\beta} F^{\alpha\beta})
+ {1\over 2} \tr (D_\alpha \phi_a D^\alpha \phi^a) - V(\phi),     \eqno(2.15)
$$
where the Higgs potential $V(\phi)$ is given by
$$
V(\phi) = - {1\over 4} \tr (F_{ab} F^{ab}).                       \eqno(2.16)
$$
The characteristic mass scale is the scale $m$ introduced in (2.2).
Since the group is $U_2$ there are four gauge bosons $A^0, A^a$. In the
`broken phase' the Higgs kinematical term yields a mass term
$$
{\cal L}_m =  {1\over 2} m^2 A^{\alpha a} A_\alpha^b \,
\tr ([\lambda_a, \lambda_c] [\lambda_b, \lambda_d]) g^{cd}.       \eqno(2.17)
$$
From (2.1) we see that the gauge bosons acquire masses given by
$$
m_0 = 0, \qquad m_a = 2 m.                                        \eqno(2.18)
$$
For more details we refer to Dubois-Violette {\it et al.} (1989).

To construct the lagrangian (2.15) we used in an essential way a metric
on the extra algebraic structure we added to space-time; it appears as
the last factor in (2.17).  We chose this metric to be the Killing
metric defined in (2.1). This is the only metric with respect to which
the derivations (2.2) are Killing derivations. We could have chosen
however another one, for example that given by
$$
\tilde g_{ab} = {\rm diag} (1, 1, \epsilon^2).                    \eqno(2.19) 
$$
We would find then the mass spectrum
$$
\tilde m_0 = 0, \qquad 
\tilde m_1 = \tilde m_2 = \sqrt 2 \epsilon^{-1} m, \qquad  
\tilde m_3 = 2 m.                                                 \eqno(2.20)
$$
The two modes $A^1$ and $A^2$ decouple in the limit 
$\epsilon \rightarrow 0$.

From the second term in the lagrangian (2.15) we see that if we use the
metric (2.19) then we must renormalize the amplitudes of the scalar
fields:
$$
\tilde \phi_1 = \phi_1, \qquad 
\tilde \phi_2 = \phi_2, \qquad
\tilde \phi_3 = \epsilon^{-1} \phi_3.                             \eqno(2.21)
$$
The non-vanishing components of $F_{ab}$ are given by
$$
\eqalign{
&\tilde F_{23} = \epsilon [\tilde \phi_2, \tilde \phi_3] 
               - m\,C^1{}_{23}\tilde \phi_1,                   \cr
&\tilde F_{31} = \epsilon [\tilde \phi_3, \tilde \phi_1] 
               - m\,C^2{}_{31} \tilde \phi_2,                  \cr
&\tilde F_{12} = [\tilde \phi_1, \tilde \phi_2] 
               - \epsilon m\,C^3{}_{12} \tilde \phi_3,
}                                                                 \eqno(2.22)
$$
and the Higgs potential is given by
$$
V(\phi) = - {1 \over 2} \tr (\epsilon^{-2}\tilde F_{23}^2 
                           + \epsilon^{-2}\tilde F_{31}^2 
                                        + \tilde F_{12}^2).       \eqno(2.23)
$$

In the `broken phase' the Higgs field $\phi$ acquires the non-vanishing
vacuum value $\phi = - \theta$ and therefore from (2.13) the modes are
described by the coefficients in the expansion of the gauge potential:
$$
\omega = (\omega^a_b \lambda_a + {i\over \sqrt 2} \omega^0_b) \theta^b.
                                                                  \eqno(2.24)
$$
If we renormalize as above and retain only quadratic terms we find
$$
V(\phi) \simeq - {1 \over 2} m^2 \tr (\epsilon^{-2} c^2_1 
                                    + \epsilon^{-2} c^2_2 
                                                  + c^2_3)        \eqno(2.25)
$$
with
$$
\eqalign{
&c_1 = (\epsilon C^a{}_{2b} \tilde \omega^b_3  
     - C^a{}_{3b} \tilde \omega^b_2 
     + C^1{}_{23} \tilde \omega^a_1) \lambda_a 
     + {i\over \sqrt 2} C^1{}_{23} \tilde \omega^0_1,    \cr
&c_2 = (C^a{}_{3b} \tilde \omega^b_1  
     - \epsilon C^a{}_{1b} \tilde \omega^b_3 
     + C^2{}_{31} \tilde \omega^a_2) \lambda_a 
     + {i\over \sqrt 2} C^2{}_{31}  \tilde \omega^0_2,    \cr
&c_3 = (C^a{}_{1b} \tilde \omega^b_2  
     - C^a{}_{2b} \tilde \omega^b_1 
     + \epsilon C^3{}_{12} \tilde \omega^a_3) \lambda_a 
     + {i\over \sqrt 2} C^3{}_{12}  \tilde \omega^0_3.
}                                                                  \eqno(2.26)
$$
That is, we have to leading order in $\epsilon$
$$
\epsilon^2 m ^{-2} V(\phi) \simeq  
  2 (\tilde \omega^1_2 + \tilde \omega^2_1)^2 
+ 2 (\tilde \omega^1_1 - \tilde \omega^2_2)^2 
+ (\tilde \omega^3_2)^2 + (\tilde \omega^3_1)^2 
+ (\tilde \omega^0_1)^2 + (\tilde \omega^0_2)^2.                   \eqno(2.27)
$$
We find then that
$$
\eqalign{
&\tilde \omega^1_2 + \tilde \omega^2_1 \rightarrow 0,   \qquad
 \tilde \omega^3_2 \rightarrow 0,                       \qquad
 \tilde \omega^0_1 \rightarrow 0,                       \cr
&\tilde \omega^1_1 - \tilde \omega^2_2 \rightarrow 0,   \qquad
 \tilde \omega^3_1 \rightarrow 0,                       \qquad 
 \tilde \omega^0_2 \rightarrow 0.                       
}                                                                  \eqno(2.28)
$$
in the limit $\epsilon \rightarrow 0$. But to within a gauge
transformation we have
$$
\tilde \omega^2_3 = \tilde \omega^3_2, \qquad
\tilde \omega^3_1 = \tilde \omega^1_3, \qquad
\tilde \omega^1_2 = \tilde \omega^2_1.                             \eqno(2.29)
$$
Therefore in the limit we have
$$
\omega \rightarrow \tilde \omega^1_1 (\lambda_1 \theta^1 + 
\lambda_2 \theta^2) + \chi_3 \theta^3                              \eqno(2.30)
$$
where we have set
$$
\chi_3 = {i \over \sqrt 2} \tilde \omega^0_3 
       + \tilde \omega^3_3 \lambda_3.                              \eqno(2.31)
$$
There remain 3 modes, a real scalar field $\omega^1_1$ and a real
scalar doublet $\chi_3$. If we impose a reality condition and
reduce the algebra $M_2$ to the algebra $\bbbh$ of quaternions the
field $\tilde \omega^0_3$ will not be present and $\chi_3$ will be a
singlet.

\section{The commutative algebra}

The algebra $M_2$ has a natural $\bbbz_2$ grading 
$M_2 = M_2^+ \oplus M_2^-$ with the unit matrix and $\lambda_3$ even
and $\lambda_1$ and $\lambda_2$ odd. Let $\eta$ be an antihermitian odd
matrix with
$$
\eta^2 = - 1                                                       \eqno(3.1)
$$
and define the differential $d$ of an arbitrary element 
$\alpha \in M_2$ by
$$
d \alpha = - [\eta, \alpha]                                        \eqno(3.2)
$$
with a graded commutator. From (3.1) it follows that
$$
d \eta + \eta^2 = 1.                                                \eqno(3.3) 
$$
The unit on the right is the unit in $M_2$ considered as a 2-form.
Equation (3.3) is to be compared with (2.12). For all $p \geq 0$ we set 
$$
\Omega^{2p}(\bbbc \times \bbbc) = M_2^+, \qquad
\Omega^{2p+1}(\bbbc \times \bbbc) = M_2^-.                          \eqno(3.4)
$$  
Then $\Omega^*(\bbbc \times \bbbc)$ is a differential calculus over the
algebra $\bbbc \times \bbbc = M_2^+$. 

One defines a Yang-Mills potential to be an anti-hermitian element
$\omega \in \Omega^1(\bbbc \times \bbbc) = M_2^-$. We can write it using
$\eta$ as
$$
\omega = \eta + \phi.                                               \eqno(3.5)
$$
The unitary elements $U_1 \times U_1$ of the algebra $M_2^+$ can be
considered as the group of gauge transformations. For 
$g \in U_1 \times U_1$ we have $\omega \mapsto g^{-1} \omega g + g^{-1} dg$.  
It is clear that in particular $\eta \mapsto \eta$ and therefore that 
$\phi \mapsto g^{-1} \phi g$. It follows from (3.3) that the field
strength $F$ of the potential $\omega$ is given by
$$
F = d\omega + \omega^2 = \phi^2 + 1.                                \eqno(3.6) 
$$

Using $\Omega^*(\bbbc \times \bbbc)$ one can construct a differential
calculus over ${\cal C}(\bbbr^4) \otimes (\bbbc \times \bbbc)$, the 
algebra of functions on a double-sheeted space-time. The electromagnetic
lagrangian in the extended space can be written
$$
{\cal L} = {1\over 4} \tr (F_{\alpha\beta} F^{\alpha\beta})
         + {1\over 2} \mu^2 \tr (D_\alpha \phi D^\alpha \phi)
         + {1\over 4} \mu^4 \tr \big((\phi^2 + 1)^2\big).           \eqno(3.7) 
$$
The parameter $\mu^{-1}$ has the dimensions of length and is a measure
of the distance between the two sheets of space-time.
Since the group is $U_1 \times U_1$ there are two gauge bosons 
$A^0, A^3$. In the `broken phase' they have masses given by
$$
m_0 = 0, \qquad m_3 = \mu.                                          \eqno(3.8)
$$
For more details we refer to Connes \& Lott (1990), or to Coquereaux {\it
et al.} (1991, 1993).  See also Dubois-Violette {\it et al.} (1991).

Contrary to (2.15) the lagrangian (3.7) does not involve a metric on the
extra algebraic structure, which does not in fact possess one. Comparing
(3.7) with the lagrangian (2.15) equipped with the metric defined by
(2.19) we find that they have the same gauge-boson spectrum in the limit
$\epsilon \rightarrow 0$ provided we set
$$
\mu = 2 m.                                                          \eqno(3.9)
$$

\section{The contraction}

Let ${\cal A}$ be an associative algebra and ${\cal A}_\epsilon$ a
1-parameter family of such algebras with ${\cal A}_0 = {\cal A}$. Then
${\cal A}_\epsilon$ is a deformation of ${\cal A}$. Deformations of
associative algebras have been studied in general by Gerstenhaber
(1964); all regular deformations of a simple algebra are trivial. We are
interested in the algebra $\bbbc \times \bbbc$ as a singular contraction
of $M_2$ and in the extension of the contraction to a map from the
differential calculus of Section~2 into that of Section~3. In defining
the contraction it is convenient to use the universal calculus.

Over any arbitrary associative algebra ${\cal A}$ one can construct
(Karoubi 1983) the universal calculus $\Omega_u^*({\cal A})$.  One
defines $\Omega_u^1({\cal A})$ to be the kernel of the multiplication
map which takes ${\cal A} \otimes_\bbbc {\cal A}$ into ${\cal A}$ and
for each $p \geq 2$ one sets
$$
\Omega_u^p({\cal A}) = \Omega_u^1({\cal A}) \otimes_{\cal A} \cdots
                       \otimes_{\cal A} \Omega_u^1({\cal A})       \eqno(4.1)
$$
where the tensor product on the right contains $p$ factors. The
differential $d_u$ which takes ${\cal A}$ into $\Omega_u^1({\cal A})$ is
given, for arbitrary $a \in {\cal A}$, by 
$$
d_u a = 1 \otimes a - a \otimes 1.                                 \eqno(4.2)
$$
It can be extended to a map of $\Omega_u^p({\cal A})$ into
$\Omega_u^{p+1}({\cal A})$ by Leibniz's rule. There is a projection
$\phi$ of $\Omega_u^*({\cal A})$ onto every other differential calculus
over ${\cal A}$ given by 
$$
\phi (d_u a) = d a.                                                \eqno(4.3)
$$  

Let $T^*$ be the tensor calculus over the vector space spanned by the
$\theta^a$. Then the universal differential calculus over $M_2$ is given
by $\Omega_u^*(M_2) = M_2 \otimes_\bbbc T^*$. Comparing this with (2.8)
we see that there is a canonical imbedding,
$$
\Omega^*(M_2) \hookrightarrow \Omega_u^*(M_2),                     \eqno(4.4)
$$ 
and that for $p = 1$ this imbedding is an isomorphism:
$$
\Omega^1(M_2) \simeq \Omega_u^1(M_2).                              \eqno(4.5)
$$ 
If we introduce the element 
$$
\theta_u =  - {1 \over 2} \lambda_a d_u \lambda^a                  \eqno(4.6)
$$ 
then the isomorphism can be written as
$$
\phi (\theta_u) = \theta.                                          \eqno(4.7) 
$$

The calculus (3.4) we constructed in Section~3 can in fact be identified
with the universal calculus over $\bbbc \times \bbbc$:
$$
\Omega^*(\bbbc \times \bbbc) \simeq \Omega^*_u(\bbbc \times \bbbc).\eqno(4.8)
$$
If we introduce the element 
$$
\eta_u =  - \lambda_3 d_u \lambda_3                                \eqno(4.9)
$$ 
then the isomorphism can be written as
$$
\phi (\eta_u) = \eta.                                             \eqno(4.10) 
$$
If we extend $\phi$ to $\Omega^2_u(\bbbc \times \bbbc)$ we find from
(3.1) that
$$
\phi \big((d_u \lambda_3)^2\big) = - 2,                           \eqno(4.11)
$$
where the right-hand side is considered as an element of 
$\Omega^2(\bbbc \times \bbbc) = M_2^+$.

From general arguments it is known that any deformation of the
algebraic structure of $M_2$ can be expressed as a deformation of the
generators.  For each $\epsilon \geq 0$ consider the change of basis
$$
\lambda^\prime_a = (\epsilon\lambda_1, \epsilon\lambda_2, \lambda_3).
                                                                  \eqno(4.12)
$$
Equation (2.1) can be rewritten in terms of the new 
basis with 
$$
C^{\prime 1}{}_{23} = C^1{}_{23}, \qquad
C^{\prime 2}{}_{31} = C^2{}_{31}, \qquad
C^{\prime 3}{}_{12} = \epsilon^2 C^3{}_{12}
$$
and
$$
g^\prime_{ab} = {\rm diag} (\epsilon^2, \epsilon^2, 1).           \eqno(4.13) 
$$
The quantity $C^\prime_{abc} = g^\prime_{ad} C^{\prime d}{}_{bc}$
remains completely antisymmetric as it must by general arguments.  Let
${\cal A}_\epsilon$ be the algebra generated by the $\lambda^\prime_a$.
Then ${\cal A}_\epsilon = M_2$ for all $\epsilon > 0$ but the sequence
has the algebra $\bbbc \times \bbbc$ as a singular limit. There are two
steps involved here. The first is the singular contraction 
$\epsilon \rightarrow 0$ which leaves two nilpotent elements
$(\lambda^\prime_1, \lambda^\prime_2)$. To obtain $\bbbc \times \bbbc$
one must quotient with respect to the ideal (the radical) generated by
these elements.

Under the deformation the metric $\tilde g_{ab}$ defined in (2.19)
becomes 
$$
\tilde g^\prime_{ab} = \epsilon^2 {\rm diag} (1, 1, 1).
$$
So the deformation of the algebra (4.12) coupled with the deformation
(2.19) yields again a diagonal metric. This establishes the relation
between the two deformations (2.19) and (4.12).

We must show that this contraction can be lifted to a contraction of 
$\Omega^*(M_2)$ into $\Omega^*(\bbbc \times \bbbc)$ which respects the
action of the differentials:
$$
\def\normalbaselines{\baselineskip=18pt}
\matrix{
\Omega^*(M_2)& \buildrel {\epsilon \rightarrow 0} \over \longrightarrow 
&\Omega^*(\bbbc \times \bbbc)                                      \cr
d \downarrow \phantom{d}&& d \downarrow \phantom{d}                \cr
\Omega^*(M_2)& \buildrel {\epsilon \rightarrow 0} \over \longrightarrow 
&\Omega^*(\bbbc \times \bbbc)
}
\def\normalbaselines{\baselineskip=12pt}                          \eqno(4.14)
$$
It is obvious that the contraction of the algebra induces a contraction
of the corresponding universal calculus,
$$
\Omega^*_u(M_2) \buildrel {\epsilon \rightarrow 0} \over \longrightarrow 
\Omega^*_u(\bbbc \times \bbbc),                                   \eqno(4.15)
$$
which respects the action of the differentials $d_u$.  We noticed also
in (4.4) that the forms over $M_2$ can be considered as elements of the
universal algebra. The extension (4.14) is therefore uniquely determined
by the original contraction of the algebra as a composition of (4.4),
(4.15) and (4.8). 

The inverse of (4.15) is an imbedding of $\Omega^*_u(\bbbc \times \bbbc)$
into $\Omega^*_u(M_2)$ which coupled with the projection of
$\Omega^*_u(M_2)$ onto $\Omega^*(M_2)$  yields a homomorphism of
differential algebras
$$
\Omega^*(\bbbc \times \bbbc) \buildrel \psi \over \rightarrow 
\Omega^*(M_2),                                                    \eqno(4.16)
$$
under which
$$
\psi(\eta) = - m (\lambda_1 \theta^1 + \lambda_2 \theta^2)        \eqno(4.17)
$$
and therefore
$$
\psi(\eta^2) = \epsilon^2 m^2 C^3{}_{12}\lambda_3 \theta^1 \theta^2
             = - m^2 \lambda_3 d\theta^3 = {1\over 2} (d\lambda_3)^2.
                                                                  \eqno(4.18)
$$
Using (4.11) we see that this is compatible with (3.1).  We can conclude
from (4.17) that $\psi$ restricted to  $\Omega^1(\bbbc \times \bbbc)$ 
and to $\Omega^2(\bbbc \times \bbbc)$ is a monomorphism and that, since 
the $\theta^a$ anticommute,
$$
\psi\big(\Omega^p(\bbbc \times \bbbc)\big) = 0, \qquad p \geq 3.  
$$
If we compare (2.10) with (4.17) we find that
$$
\theta = \psi(\eta) - m \lambda_3 \theta^3.                       \eqno(4.19)
$$
The second term commutes with the elements of $M_2^+$ and so (4.19) is
compatible with (2.11) and (3.2). One sees also that because of (4.18), 
(2.12) is compatible with (3.3).

One can use the homomorphism $\psi$ to construct a new differential
algebra $\Omega^{*\prime}(\bbbc \times \bbbc)$ over 
$\bbbc \times \bbbc$ given by
$$
\eqalign
{
&\Omega^{p\prime}(\bbbc \times \bbbc) = 
\Omega^p(\bbbc \times \bbbc),                 \quad p \leq 2, \cr
&\Omega^{p\prime}(\bbbc \times \bbbc) = 0     \quad p \geq 3,
}                                                                 \eqno(4.20)
$$
which one can then extend to an algebra 
$\Omega^{*\prime\prime}(\bbbc \times \bbbc)$ by adding a 1-form $\eta^3$
with the relations
$$
\lambda_3 \eta^3 = \eta^3 \lambda_3,   \qquad
\eta^3 \eta = - \eta \eta^3,           \qquad
(\eta^3)^2 = 0,                        \qquad
d\eta^3 = - 2 \lambda_3.
$$
Notice that if $\lambda_3$ is considered as a 2-form then $d\lambda_3 = 0$ 
in $\Omega^{*\prime}(\bbbc \times \bbbc)$. The differential algebra
$\Omega^{*\prime\prime}(\bbbc \times \bbbc)$ is more similar to that
introduced by Connes \& Lott over the algebra $\bbbh \times \bbbc$ than
to the original $\Omega^*(\bbbc \times \bbbc)$.  The homomorphism $\psi$
can be extended to $\Omega^{*\prime\prime}(\bbbc \times \bbbc)$ by
setting
$$
\psi (\eta^3) = m \theta^3                                        \eqno(4.21)
$$
and (4.19) can be rewritten as 
$$
\theta = \psi(\eta - \lambda_3 \eta^3).
$$

The contraction of the potential $\omega$ and the associated field
strength $F$ of Section~2 onto those of Section~3 is in principle 
uniquely and well defined. However under the deformation each individual
mode has to be renormalized so that the coefficient in the kinetic term
remains constant. In other words the coefficient of the modes has to be
written using the normalized basis $\theta^a$ of the 1-forms and these
do not vanish under the contraction. In fact $\theta^1$ and $\theta^2$
are singular and $\theta^3$ has a nonvanishing limit.  From (4.17) we
see that (2.30) can be written as
$$
m \omega \rightarrow  m \chi_3 \theta^3 - \tilde \omega^1_1 \psi(\eta). 
                                                                  \eqno(4.22)
$$
Although $\chi_3$ is a function with values in $\bbbc \times \bbbc$ the
second term of the right-hand side of (4.20) is not the image of an
element of $\Omega^1(\bbbc \times \bbbc)$ under $\psi$.  The image of
the 2-point model of Connes \& Lott and Coquereaux under the homomorphism 
$\psi$ is found to be equal to the singular deformation (2.19) of the
model of Dubois-Violette {\it et al.} with the addition of a real scalar
gauge-invariant doublet $\chi_3$. If one uses the differential calculus
$\Omega^{*\prime\prime}(\bbbc \times \bbbc)$ one can rewrite (4.22) as
$$
m \omega \rightarrow \psi(\chi_3 \eta^3 - \tilde \omega^1_1 \eta).
$$
The 2-point model with the new differential calculus is identical to the
singular deformation (2.19) of the model of Dubois-Violette {\it et
al.}

\section{Conclusions} 

We have shown that the 1-forms of the universal differential calculus
over $\bbbc \times \bbbc$ can be considered as a singular limit of a
sequence of 1-forms of a differential calculus which is `smooth' in the
sense that it is based on the derivations of an algebra.  Equivalently
we have shown that to within a real scalar doublet the 2-point model of
Connes \& Lott and Coquereaux is a singular contraction of a model defined
on a `smooth' geometry. An obvious extension would be to investigate to
what extent the extended model of Connes \& Lott, using the algebra 
$\bbbh \times \bbbc$, can be obtained as a singular limit of geometries
which are `smooth'.

\section*{Acknowledgment}

This research was partially supported by the Franco-Polish Research
Program No. 5255, financed by the French Embassy in Poland and the KBN.
One of the authors (J.M.) would like to thank A. Chakrabarti, M.
Dubois-Violette, T. Masson and H. R\"omer for interesting conversations.
\parskip 2pt plus 1pt 
\parindent=0cm

\section*{References}

Bratteli O., Robinson D.W. 1979, {\it Operator Algebras and Quantum 
Statistical Mechanics I}, Springer Verlag.

Connes A. 1994, {\it Noncommutative Geometry}, Academic Press.

--- 1995, {\it Noncommutative geometry and reality}, 
J. Math. Phys. {\bf 36} 6194.

Connes A., Lott J. 1990, {\it Particle Models and Noncommutative Geometry},
in `Recent Advances in Field Theory', Nucl. Phys. Proc. Suppl. {\bf B18} 29.

--- 1992, {\it The metric aspect of non-commutative geometry},
Proceedings of the 1991 Carg\`ese Summer School, Plenum Press.

Coquereaux R., Esposito-Far\`ese G., Vaillant G. 1991, {\it Higgs Fields as
Yang-Mills Fields and Discrete Symmetries}, Nucl. Phys. {\bf B353} 689.

Coquereaux R., H\"au\ss ling R., Scheck F. 1993, {\it Algebraic
Connections on Parallel Universes}, Int. J. Mod. Phys. {\bf A10} 89.

Deser S. 1957, {\it General Relativity and the Divergence Problem in
Quantum Field Theory}, Rev. Mod. Phys. {\bf 29} 417.

Dubois-Violette M. 1988, {\it D\'erivations et calcul diff\'erentiel 
non-commutatif}, C. R. Acad. Sci. Paris {\bf 307} S\'erie I 403.

Dubois-Violette M., Kerner R., Madore J. 1989, {\it Gauge bosons in a
noncommutative geometry}, Phys. Lett. {\bf B217} 485; {\it Classical
bosons in a noncommutative geometry}, Class. Quant. Grav. {\bf 6} 1709.

--- 1991, {\it Super Matrix Geometry}, Class. Quant. Grav. {\bf 8} 1077.

Dubois-Violette M., Madore J., Masson T., Mourad J. 1995, {\it On 
Curvature in Noncommutative Geometry}, Preprint LPTHE Orsay 95/63.

Gerstenhaber M. 1964, {\it On the Deformation of Rings and Algebras},
Ann. Math. {\bf 78} 59.

Iochum B., Sch\"ucker T. 1994, {\it A Left-Right Symmetric Model
\`a la Connes-Lott}, Lett. Math. Phys. {\bf 32} 153.

Isham C.J., Salam A., Strathdee J. 1971, {\it Infinity Suppression in 
Gravity-Modified Quantum Electrodynamics}, Phys. Rev. {\bf D3} 1805.

Karoubi J. 1983, {\it Homologie cyclique des groupes et des alg\`ebres},
C. R. Acad. Sc. Paris {\bf 297} 381.

Madore J. 1990, {\it Modification of Kaluza-Klein Theory}, 
Phys. Rev. {\bf D41} 3709.

--- 1995, {\it An Introduction to Noncommutative Differential Geometry
and its Physical Applications}, Cambridge University Press.

Madore J., Mourad J. 1995, {\it On the Origin of Kaluza-Klein Structure}, 
Phys. Lett. {\bf B359}, 43.

\end{document}